\PassOptionsToPackage{numbers,sort&compress}{natbib}
\documentclass[a4paper,fleqn]{cas-sc}

\usepackage{amsmath}
\usepackage{iftex}
\ifPDFTeX
  \usepackage[utf8]{inputenc}
  \usepackage[T1]{fontenc}
  \IfFileExists{stix.sty}{}{\IfFileExists{lmodern.sty}{\usepackage{lmodern}}{\typeout{*** lmodern.sty not found: install lmodern or cm-super for vector fonts ***}}}
\fi
\usepackage{graphicx}
\usepackage{booktabs}
\graphicspath{{figures/}}

\usepackage{longtable}
\usepackage{array}
\usepackage{calc}

\begin{document}
\let\WriteBookmarks\relax
\def\floatpagepagefraction{1}
\def\textpagefraction{.001}

\shorttitle{Grain-boundary segregation and constant-stress flow in nanocrystalline Fe--18Cr--12Ni}
\shortauthors{A. Kumar}

\title[mode=title]{Grain-boundary segregation delays the onset of plastic flow in nanocrystalline Fe--18Cr--12Ni}

\author[1]{Ashwinee Kumar}[orcid=0000-0003-0384-5108]
\cormark[1]
\ead{ashwinee.met@nitjsr.ac.in}
\credit{Conceptualization, Methodology, Software, Formal analysis, Investigation, Data curation, Writing -- original draft, Writing -- review and editing, Visualization}
\affiliation[1]{organization={Department of Metallurgical and Materials Engineering, National Institute of Technology Jamshedpur},
            city={Jamshedpur},
            postcode={831014},
            state={Jharkhand},
            country={India}}
\cortext[1]{Corresponding author.}

\begin{abstract}
Grain-boundary (GB) segregation in austenitic Fe--Cr--Ni alloys has been studied mainly in irradiated
or sensitised material, rarely in the nanocrystalline state. Hybrid molecular dynamics/Monte
Carlo (MD/MC) simulations are used to segregate three independent nanocrystalline Fe--18Cr--12Ni
polycrystals, which are compared with annealed and random controls on the same microstructures. Ni is depleted at the GB plane and Cr is enriched in the
adjacent atomic shell; both directions are established at the production sampling budget and
strengthen when one of the three microstructures is sampled further. The Cr enrichment is resolved only against the far-field grain interior; the common partition into
face-centred-cubic (fcc) and non-fcc atoms places the enriched shell in its own reference. Under a constant uniaxial
stress of 4.6 GPa at 300 K, the segregated polycrystals take about 16\% longer than random ones to
reach 50\% strain; measured against annealed controls, boundary chemistry contributes about two-thirds
of this gain and boundary relaxation the rest. The difference arises from a lower strain rate early in plastic flow, with no resolvable
change in dislocation content. At a fixed strain rate the segregated state has a 6--7\% higher peak stress than its annealed
control at 300, 750 and 950 K, and at matched ratios of applied to flow stress the gain is the same
at all three temperatures, decreasing only as that ratio rises. Early plastic flow is thus the stage at which GB chemistry acts, and simulated GB composition in
nanocrystals depends strongly on the reference region.
\end{abstract}


\maketitle
\section{Introduction}\label{introduction}

Nanocrystalline metals, with grain sizes below about 100 nm, owe much of their strength to the high
density of grain boundaries, and at grain sizes of 10--20 nm a substantial fraction of all atoms lies
in or next to a boundary \cite{ref1,ref2}. The same boundaries limit their usefulness at elevated
temperature: grain growth, boundary sliding and boundary-mediated plasticity lower the resistance to
time-dependent deformation, and the creep rates of nanocrystalline metals can exceed those of their
coarse-grained counterparts by orders of magnitude \cite{ref43,ref52}. Stabilising the boundaries
chemically, so that they neither migrate nor slide easily, has produced nanocrystalline alloys with
creep resistance far above that of the pure metal \cite{ref52}.

Solute segregation to grain boundaries is the principal route to such stabilisation. It occurs when a
solute lowers the interfacial free energy, with an equilibrium coverage set by the segregation energy
\cite{ref3,ref4,ref5}. In coarse-grained alloys segregation is associated mainly with embrittlement
\cite{ref6}. In nanocrystalline alloys it can reduce the boundary energy enough to suppress grain growth
\cite{ref10,ref33,ref34}, raise the strength through boundary doping \cite{ref36,ref37}, and, at higher
solute levels, produce boundary complexions and amorphous intergranular films that change the
deformation response \cite{ref38,ref39,ref40}. Relaxation of the boundary structure by annealing,
without any change in chemistry, also strengthens nanocrystalline alloys \cite{ref50}, so the chemical
and structural contributions of a heat treatment must be separated before either can be attributed.

Austenitic Fe--Cr--Ni stainless steels are among the most widely used structural alloys, and their
grain-boundary chemistry has been studied mainly in two non-equilibrium situations. Thermal
sensitisation depletes the boundaries in chromium by precipitation of chromium-rich M$_{23}$C$_6$
carbides, and radiation-induced segregation depletes chromium and enriches nickel through the coupling
of solute fluxes to a sustained point-defect flux \cite{ref7,ref8,ref9,ref41}. In ion-irradiated
austenitic stainless steel, the grain size and ion implantation both affect the resulting boundary
segregation \cite{ref53}. Chromium segregation at the boundaries of a 316 stainless steel has
been related to the boundary structure \cite{ref54}, and atom-probe studies resolve boundary segregation in
Ni--Cr--Fe alloys before carbide nucleation and at austenite grain boundaries of steels
\cite{ref27,ref28}.

Atomistic simulation gives access to the segregation tendency of the alloy itself. Hybrid molecular
dynamics/Monte Carlo schemes, which interleave substitutional swaps with dynamical relaxation, sample
the species distribution at fixed composition without imposing a profile \cite{ref12,ref11}. In
nanocrystalline Fe(Cr), such simulations have been compared with the boundary-type dependence of
chromium segregation measured by atom-probe tomography \cite{ref55}; in fcc multiprincipal element
alloys, they reveal an enrichment of Fe in the fcc region adjacent to the boundary \cite{ref56}. Hybrid MC/MD segregation followed by deformation of the segregated bicrystals has been
used to relate boundary chemistry to the shear response of CoCrFeNi-based high-entropy alloys
\cite{ref57}. For the Fe--Ni--Cr system, high-throughput calculations predict chromium enrichment and
nickel depletion at the boundaries at intermediate temperatures \cite{ref26}, first-principles
calculations give segregation energies of transition-metal solutes at \(\gamma\)-Fe boundaries
\cite{ref25}, and the distribution of segregation energies over the sites of a polycrystal can be
computed directly \cite{ref31,ref32}. Molecular-dynamics studies of the mechanical consequences in
nanocrystalline FeCrNi have, however, imposed the boundary composition rather than sampled it, in one
recent case as a nickel enrichment of the boundaries \cite{ref58}.

Molecular dynamics has also been used to study time-dependent deformation of nanocrystalline metals
under constant stress, from grain-boundary diffusion creep in palladium \cite{ref59} to alloyed
nanocrystalline magnesium \cite{ref60}. In nanocrystalline Ni, segregation of Mo to the boundaries
lowers the simulated creep rate \cite{ref61}, in line with the experimental creep
resistance of chemically stabilised nanocrystalline alloys \cite{ref52}. These studies analyse mainly steady-state rates and
their dependence on stress and grain size; the stage of the deformation history at which boundary chemistry acts, and the separation
of its effect from that of boundary relaxation, have received less attention.

Two further questions arise specifically in the nanocrystalline regime. When 30\% of the atoms are
classified as boundary atoms, the choice of the interior reference used to express boundary
composition is no longer innocuous, because much of the nominal interior lies within a few ångström
of a boundary. And when segregation is sampled rather than imposed, the extent to which the sampled
state has converged determines how its magnitudes, as distinct from its directions, can be used.

The objective of the present work is to determine the grain-boundary segregation tendency of
nanocrystalline Fe--18Cr--12Ni by hybrid MD/MC simulation, with explicit attention to the reference
used to measure boundary composition and to the convergence of the sampling, and to identify how the
resulting boundary chemistry changes the response of the same microstructures to constant uniaxial
stress. Three independent polycrystals are compared with annealed and random controls on the same
tessellations, the deformation response is resolved by strain, the applied stress is referred to the
flow stress measured at constant strain rate, and the defect structure is compared at matched strain.

\section{Methodology}\label{methodology}

\subsection{Nanocrystalline model construction}\label{nanocrystalline-model-construction}

\begin{figure}[pos=htbp]\centering
  \includegraphics{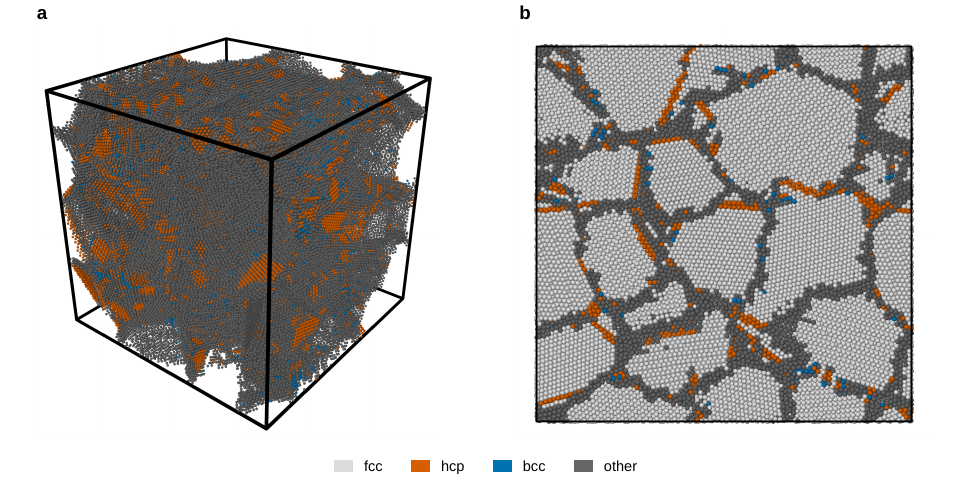}
  \caption{Nanocrystalline Fe--18Cr--12Ni model (microstructure 1, segregated state, 300~K, before
  loading), coloured by adaptive common-neighbour analysis. (a)~The grain-boundary network, with
  fcc atoms hidden; hcp atoms mark stacking faults and twin boundaries. (b)~A 1.5~nm thick slab
  through the mid-plane of the $\approx 20$~nm periodic cell, showing the grains ($\approx 7$~nm
  mean size) and the boundary network that separates them.}
  \label{fig:model}
\end{figure}

All simulations use a nanocrystalline Fe--18Cr--12Ni austenitic model alloy (Fe70/Cr18/Ni12 at.\%),
constructed as a periodic Voronoi polycrystal in the face-centred-cubic phase with approximately 40
grains of mean size \(\approx 7\) nm in a \(\approx 200\) Å periodic cube (Fig.~\ref{fig:model}).
Three statistically independent tessellations were generated from different random seeds and are
referred to throughout as microstructures 1, 2 and 3. Atoms closer than 2 Å across the
grain-boundary seams of the initial Voronoi packing (\(\approx 4.2\%\) of the sites; 27,450 of
654,085 for microstructure 1) were removed before any dynamics, leaving 626,635, 626,612 and
627,051 atoms respectively. Because the three tessellations differ in grain topology,
grain-boundary area and boundary character distribution, they constitute independent
microstructural realisations rather than repeated sampling of one structure, and all statistics
reported in this work are computed across them.

Lattice sites were decorated randomly at the target composition to produce the random
reference state. Each microstructure was then relaxed by conjugate-gradient energy minimisation and
equilibrated in the isothermal--isobaric ensemble at zero pressure. Tessellations were built with
Atomsk \cite{ref14}, all dynamics were run in LAMMPS \cite{ref16} (version 22 Jul 2025, Update 4,
KOKKOS accelerator package), and structural analysis and visualisation were performed in OVITO
\cite{ref19} (version 3.16.1). The simulation parameters are collected in Table~\ref{tab:params}.

\begin{table}[pos=htbp]
\caption{Simulation parameters.}\label{tab:params}
\centering\small
\begin{tabular}{@{}p{0.30\linewidth}p{0.64\linewidth}@{}}
\toprule
parameter & value \\
\midrule
composition & Fe--18Cr--12Ni (Fe70/Cr18/Ni12 at.\%) \\
microstructure & periodic Voronoi polycrystal, $\approx$40 grains, mean size $\approx$7 nm \\
cell & $\approx$200 Å periodic cube; 626,635 / 626,612 / 627,051 atoms (microstructures 1/2/3) \\
interatomic potential & EAM, Bonny et al.\ Fe--Ni--Cr \cite{ref15} \\
MD time step & 2 fs, velocity-Verlet integration \\
thermostat / barostat & Nos\'e--Hoover; \(T_\mathrm{damp} = 0.2\) ps, \(P_\mathrm{damp} = 5.0\) ps \\
segregation & hybrid MD/MC, canonical swaps at fixed composition, \(T_\mathrm{MC}\) = 900 K \\
MC swap protocol & 200 attempts per channel every 100 MD steps (0.2 ps); channels Fe--Ni, Fe--Cr, Ni--Cr; 100{,}000 MD steps (200 ps); 0.96 attempted and 0.43 accepted swaps per atom; acceptance 0.44 (Fe--Ni), 0.46 (Fe--Cr), 0.44 (Ni--Cr) \\
controls & annealed-random (identical thermal history, swaps disabled) and as-decorated random \\
equilibration before loading & 200 ps at the test temperature and zero pressure \\
loading & uniaxial constant stress, NPT, transverse directions at zero pressure \\
reference condition & 300 K, 4.6 GPa \\
condition series & 300 K/4.6 GPa, 750 K/4.0 GPa, 950 K/3.5 GPa \\
stress series & 300 K at 4.4, 4.6, 4.8 GPa \\
matched series & 750 and 950 K at \(\sigma/\sigma_\mathrm{flow}\) = 1.17, 1.22, 1.27 (3.04--3.32 and 2.52--2.74 GPa) \\
constant-strain-rate tension & \(10^{9}\) s\(^{-1}\) to \(\varepsilon = 0.20\) at 300, 750 and 950 K, all three states \\
life criterion & time to engineering strain 0.5, \(t_{0.5}\) (interpolated); runs continued to \(\varepsilon \approx 0.8\) \\
structural analysis & adaptive CNA \cite{ref44} and DXA \cite{ref45}; composition on fixed-box quenched configurations \\
replication & \(n = 3\) independent tessellations; all comparisons paired within microstructure \\
\bottomrule
\end{tabular}
\end{table}

\subsection{Interatomic potential and its validation}\label{interatomic-potential}

Atomic interactions are described by the embedded-atom-method (EAM) potential of Bonny et al.\
\cite{ref15} for the ternary Fe--Ni--Cr system, which was developed to study plasticity in
austenitic stainless steels. It treats the three species on an equal footing and supports the
substitutional exchanges required by the Monte Carlo step described below. Because it was fitted to
defect and plasticity properties rather than to grain-boundary energetics, its bulk properties at
the present composition were compared with experimental data for austenitic stainless steels of
similar composition (Table~\ref{tab:validation}). The 0 K lattice parameter was obtained by
cell relaxation of a 4000-atom random cell, and the cubic elastic constants from \(\pm 0.5\)\% strains
with relaxed internal coordinates, averaged over six random 256-atom decorations at the exact
composition. The intrinsic stacking-fault energy was taken as the mean over twelve random
decorations of the generalised stacking-fault curve.

\begin{table}[pos=htbp]
\caption{Properties of the Bonny et al.\ potential \cite{ref15} at Fe70Cr18Ni12 compared with experimental
data for austenitic stainless steels. Potential values are 0 K; the elastic constants are mean
$\pm$ SD over six random decorations and the stacking-fault energy is the mean $\pm$ 95\% confidence
interval over twelve. Experimental values are at room temperature.}\label{tab:validation}
\centering\small
\begin{tabular}{@{}lccl@{}}
\toprule
property & potential & experiment & experimental source \\
\midrule
lattice parameter \(a_0\) (Å) & 3.512 & 3.558 & 304L, neutron diffraction \cite{ref46} \\
\(C_{11}\) (GPa) & \(221.1 \pm 2.3\) & 216 & 304L, single-crystal constants \cite{ref46} \\
\(C_{12}\) (GPa) & \(139.8 \pm 1.4\) & 126 & \cite{ref46} \\
\(C_{44}\) (GPa) & \(131.3 \pm 0.5\) & 173 & \cite{ref46} \\
Zener ratio \(2C_{44}/(C_{11}-C_{12})\) & \(3.23 \pm 0.04\) & 3.84 & computed from \cite{ref46} \\
intrinsic stacking-fault energy (mJ m\(^{-2}\)) & \(16.6 \pm 2.2\) & 32.8 (12.9--42) & 316L; mean and literature range \cite{ref47} \\
\bottomrule
\end{tabular}
\end{table}

The lattice parameter is 1.3\% below the room-temperature experimental value, part of which is
thermal expansion, and \(C_{11}\) agrees within 3\%. \(C_{12}\) is 11\% too high and \(C_{44}\) 24\%
too low, so the potential underestimates the shear stiffness and the elastic anisotropy. The
intrinsic stacking-fault energy is half the mean value reported for 316L but lies inside the range
of published measurements for that steel; it places the model in the low-stacking-fault-energy regime
in which partial-dislocation slip and faulting dominate plasticity, which is the regime of the
austenitic steels it represents. Because the value lies at the low end of that range, partial slip
and faulting may be more prominent than in the real alloy; this affects both states of each pair
alike, although the relative weight of boundary and intragranular contributions may differ. The potential was not fitted to grain-boundary energies or to
solute--boundary interactions, and no first-principles segregation energies at comparable
boundaries were available for a direct test; the segregation results below are therefore the
segregation tendency of this interatomic model, compared with independent first-principles and
experimental literature in Section~\ref{sec:disc-seg}.

\subsection{Hybrid molecular-dynamics / Monte Carlo segregation}\label{hybrid-molecular-dynamics-monte-carlo-segregation}

Substitutional segregation was driven by interleaved atom-swap Monte Carlo and molecular-dynamics
relaxation, in the manner of Sadigh et al.\ \cite{ref12,ref11}. Unlike the variance-constrained
semi-grand-canonical algorithm of that work, the present scheme is canonical: swaps were attempted
between all three species pairs (Fe--Ni, Fe--Cr, Ni--Cr) at fixed composition, with candidates drawn
uniformly over the whole cell, so the alloy content is conserved exactly and only the spatial
distribution of species evolves. Trial swaps are accepted with the Metropolis criterion
\[
P_\text{acc} = \min\left[1,\ \exp\!\left(-\frac{\Delta E}{k_\text{B}T_\text{MC}}\right)\right],
\tag{1}
\]
where \(\Delta E\) is the potential-energy change of the swap and \(T_\text{MC}\) = 900 K the Monte
Carlo temperature. Molecular dynamics between swap batches allows the structure to relax locally, so
the sampled states are not restricted to the rigid lattice.

Two control states were generated for every microstructure. The annealed-random control
receives an identical thermal history, that is, the same molecular-dynamics protocol at the same
temperature for the same duration, with the swap moves disabled, and therefore isolates thermal
annealing and structural relaxation of the boundary network from chemical redistribution. The
random state is the as-decorated structure before any anneal. Segregated against
annealed-random measures the chemistry contribution, annealed-random against random the
annealing contribution, and segregated against random the total effect. Each control was
constructed on its own microstructure's tessellation. Because overlapping atoms are removed when each structure file is read,
the paired segregated and annealed structures differ by at most three atoms, and the three states of a
microstructure by at most five (8 ppm).

\begin{figure}[pos=htbp]\centering
  \includegraphics{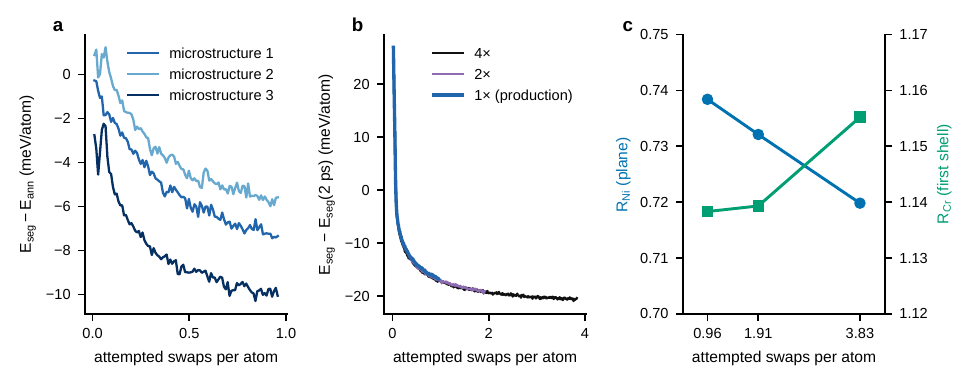}
  \caption{Monte Carlo sampling at $T_\mathrm{MC} = 900$~K. (a)~Potential energy of the segregated
  state relative to its annealed-random control against attempted swaps per atom, for the three
  microstructures at the production budget. (b)~Potential energy of the segregated state of
  microstructure 1, relative to its value at 2~ps, for one, two and four times the production budget
  from the same seeds. (c)~Boundary-plane Ni ratio and first-shell Cr ratio (Eq.~(2), quenched
  configurations) at the end of the three runs of (b).}
  \label{fig:mc}
\end{figure}

The production budget corresponds to 0.96 attempted and
0.43 accepted swaps per atom. The acceptance ratio is 0.44 for Fe--Ni and Ni--Cr exchanges and 0.46 for
Fe--Cr exchanges, with a spread below 0.005 between microstructures. The convergence
of the segregated state was tested by repeating the procedure for microstructure 1 at two and four
times this budget from the same seeds (Fig.~\ref{fig:mc}). The energy of the segregated state
relative to its annealed control is still decreasing at the end of the production run in all three
microstructures (Fig.~\ref{fig:mc}a), and the fourfold run lowers the energy by a further
3.3 meV/atom, after which its drift has fallen to 0.2 meV/atom per attempted swap per atom
(Fig.~\ref{fig:mc}b). Across the same ladder the boundary-plane nickel ratio deepens monotonically
(\(R_\text{Ni} = 0.738 \to 0.732 \to 0.720\)) and the first-shell chromium ratio rises
(\(R_\text{Cr} = 1.138 \to 1.139 \to 1.155\)), while the boundary fraction is unchanged
(0.28--0.29; Fig.~\ref{fig:mc}c and Supplementary Fig.~S1). The production state is therefore a
partially equilibrated, MC-annealed state rather than an equilibrium one: the directions of both
segregation channels are established at the production budget and are unchanged by further
sampling, whereas their magnitudes are lower bounds on the converged values.

\subsection{Grain-boundary definition and the composition estimator}\label{grain-boundary-definition-and-the-composition-estimator}

Grain boundaries were identified structurally, as the set of atoms whose local environment is not
classified as face-centred cubic by adaptive common-neighbour analysis (a-CNA) \cite{ref17,ref44}.
The same classification supplies a distance coordinate: for each atom, \(d\) is the distance to the
nearest non-fcc atom under periodic boundary conditions. Atoms with \(d = 0\) constitute the
boundary plane itself; the first fcc neighbour shell occupies \(0.5 < d < 3.08\) Å, the upper limit
being the first minimum of the \(d\) histogram. Composition at a location \(\ell\) is reported as a
ratio to the far-field grain interior,
\[
R_X(\ell) = \frac{c_X(\ell)}{c_X(d > 5\ \text{Å})},
\tag{2}
\]
with \(\ell\) either the boundary plane or the first fcc shell, so that \(R_X < 1\) denotes depletion
and \(R_X > 1\) enrichment.

The choice of reference matters. Partitioning the system into its non-fcc and fcc populations and
referencing the first to the second (the direct reading of a structural boundary definition)
fails when the non-fcc fraction is close to 0.3, as it is here: 44\% of the fcc population lies
within 5 Å of a boundary and 27\% within the first neighbour shell. A species that segregates to
those shells then enters its own reference, and the partition ratio subtracts the signal from
itself. All compositional results are therefore reported through Eq.~(2). Both cuts in Eq.~(2) were
varied, the reference over \(d > 4\)--8 Å and the shell window over 2.90--3.30 Å; every contrast
reported below keeps its sign in all three microstructures at all fifteen combinations, and its
magnitude varies by less than the between-microstructure scatter (Supplementary Table~S1).

Configurations written directly from finite-temperature dynamics are unsuitable for this
estimator: thermal displacement causes CNA to classify enough fcc interior atoms as non-fcc that
roughly 22\% of the apparent boundary population of a 900 K configuration is grain interior.
All configurations used for composition analysis were therefore quenched by fixed-box
conjugate-gradient minimisation (force tolerance \(10^{-4}\) eV/Å, at most 500 iterations) before
classification, which removes the thermal contribution while leaving the boundary network intact
(Supplementary Section~S8).

A second effect survives the quench. Measured on the annealed-random control, which is chemically
homogeneous by construction, the boundary-plane nickel ratio is 0.95 rather than unity, and the
chromium and iron ratios are 0.98 and 1.01. A label-permutation null test, in which species labels
are shuffled at fixed atomic positions and the analysis repeated (20 permutations), returns
\(1.00 \pm 0.01\) for every species at both locations. The estimator is thus unbiased, and the offset
is a composition--classification coupling: nickel stabilises the fcc environment, so nickel-poor
regions are preferentially classified as non-fcc irrespective of whether they are boundaries.
Because the offset amounts to a quarter to a half of the segregation signal, depending on species
and location, all segregation results are reported as the segregated-minus-annealed contrast
\[
\Delta R_X(\ell) = R_X^\text{seg}(\ell) - R_X^\text{ann}(\ell).
\tag{3}
\]

\subsection{Constant-stress deformation protocol}\label{sec:protocol}

Each segregated or annealed structure was carried into the mechanical test by re-drawing velocities
at the test temperature and equilibrating for 200 ps at zero pressure; Monte Carlo swaps were
disabled throughout, so the species distribution is fixed while the structure deforms. A uniaxial
tensile stress was then applied along one cube axis in the isothermal--isobaric ensemble, with the
two transverse directions held at zero pressure. The reference condition is 300 K and 4.6 GPa.
Engineering strain \(\varepsilon\) along the loading axis was recorded every 0.2 ps, and the strain
rate is
\[
\dot{\varepsilon} = \frac{\mathrm{d}\varepsilon}{\mathrm{d}t}.
\tag{4}
\]
The life metric is \(t_{0.5}\), the time at which \(\varepsilon\) first reaches 0.5, obtained by
interpolation of the strain history; runs were continued to \(\varepsilon \approx 0.8\) so that
\(t_{0.5}\) is always an interpolated value. The endpoint is a fixed strain, not fracture.

The applied stresses are far above engineering levels and the accessible time is of order
\(10^2\) ps, giving mean strain rates of order \(10^{9}\)--\(10^{10}\) s\(^{-1}\). To place the applied
stress relative to the strength of the model, all three states of every microstructure were also
deformed in uniaxial tension at a constant engineering strain rate of \(10^{9}\) s\(^{-1}\) to
\(\varepsilon = 0.20\), with the transverse directions at zero pressure and the same thermostat and
barostat settings. At 300 K these tests start from the equilibrated configurations of the
constant-stress runs; at 750 and 950 K the same structures were equilibrated for 200 ps at the test
temperature, with the random state re-decorated from the production seeds. Three quantities are taken from each stress--strain curve: the peak stress
\(\sigma_\text{peak}\); the flow stress \(\sigma_\text{flow}\), defined as the mean stress over
\(\varepsilon = 0.10\)--0.20; and the elastic modulus, from a linear fit over \(\varepsilon \le 0.01\).
All flow stresses refer to this single strain rate; their rate dependence was not measured.
At 300 K the annealed controls have \(\sigma_\text{flow} = 3.77 \pm 0.08\) GPa, so 4.6 GPa is
\(1.22\,\sigma_\text{flow}\) (1.18--1.22 across the three states) and exceeds \(\sigma_\text{peak}\) in all nine
runs (Section~\ref{sec:csr}).
With a homologous temperature of \(\approx 0.18\), the constant-stress tests therefore probe
stress-controlled plastic flow of the nanocrystal above its flow stress rather than creep in the
classical sense of sub-yield, thermally activated deformation. The descriptors conventionally used
for constant-stress tests, namely the minimum strain rate, stress exponents and the Monkman--Grant
product \cite{ref42,ref43}, are used below descriptively. No extrapolation to service conditions is
intended; the comparison that the protocol supports is between chemically distinct states of the
same microstructure under identical loading.

\subsection{Strain-rate band decomposition}\label{sec:bands}

To locate where in the deformation history a difference in \(t_{0.5}\) originates, each strain
history was decomposed by strain rather than by time. The interval \(0.03 \le \varepsilon \le 0.50\)
was divided into seven bands (0.03--0.06, 0.06--0.10, 0.10--0.15, 0.15--0.20, 0.20--0.30, 0.30--0.40,
0.40--0.50). Within each band \(b\) the mean strain rate is
\[
\dot{\varepsilon}_b = \frac{\Delta\varepsilon_b}{\Delta t_b},
\tag{5}
\]
evaluated on a Savitzky--Golay-smoothed strain history \cite{ref48} (21-point window on a 0.2 ps
grid, cubic polynomial). The quantity compared between arms is the band-rate ratio
\(\dot{\varepsilon}_b^\text{seg}/\dot{\varepsilon}_b^\text{ann}\). The first-band suppression is
\[
s_1 = 1 - \frac{\dot{\varepsilon}_{0.03-0.06}^\text{seg}}{\dot{\varepsilon}_{0.03-0.06}^\text{ann}},
\tag{6}
\]
and the integrated suppression \(s_\text{int}\) is the strain-weighted mean of the same fractional
suppression over \(0.03 \le \varepsilon \le 0.20\). The minimum strain rate
\(\dot{\varepsilon}_\text{min}\) and the Monkman--Grant product \(\dot{\varepsilon}_\text{min}\,t_{0.5}\)
\cite{ref42} are also reported. Because the endpoint is a fixed strain, this product equals
\(0.5\,\dot{\varepsilon}_\text{min}/\langle\dot{\varepsilon}\rangle\), where
\(\langle\dot{\varepsilon}\rangle = 0.5/t_{0.5}\) is the time-averaged rate; it is therefore a measure of
the shape of the strain history rather than a rupture criterion.

\subsection{Condition, stress and dose series}\label{condition-stress-and-dose-series}

The condition series compares 300 K/4.6 GPa, 750 K/4.0 GPa and 950 K/3.5 GPa. These
conditions were chosen to give comparable values of \(t_{0.5}\), so temperature and stress co-vary.
The stress series holds the temperature at 300 K and varies the applied stress over 4.4,
4.6 and 4.8 GPa, which is 1.17, 1.22 and 1.27 times the annealed flow stress at 300 K. The matched
series repeats these three ratios at 750 and 950 K, with \(\sigma_\text{flow}\) the mean annealed
flow stress at the test temperature (Section~\ref{sec:protocol}): 3.04, 3.18 and 3.32 GPa at 750 K
and 2.52, 2.63 and 2.74 GPa at 950 K. Its runs start from the equilibrated configurations of the
condition series at the same temperature, so only the applied stress differs between the two. The dose series varies the Monte Carlo temperature over 900, 700 and 500 K
at fixed 300 K/4.6 GPa test conditions, for microstructure 1 only, and is reported in
Supplementary Section~S2. The 300 K/4.6 GPa segregated runs used in Section~\ref{sec:life} were
re-segregated under a uniform protocol; those of the condition, stress and matched series come from an earlier
segregation of the same microstructures with the same parameters, and reproduce the re-segregated
\(t_{0.5}\) to within 2.5\%. Within every paired comparison both states come from the same set
(Supplementary Table~S2). The constant-strain-rate tests use the structures of
Section~\ref{sec:life} at all three temperatures; they enter the matched series only through the
annealed mean flow stress that sets its stress levels.
Stress sensitivity was characterised by a life exponent, \(t_{0.5} \propto \sigma^{-m}\), and by a
rate exponent, \(\dot{\varepsilon}_\text{min} \propto \sigma^{n}\), fitted separately for each
microstructure and arm over the three stresses; the 9\% stress lever limits both to descriptive use.

\subsection{Structural analysis during deformation}\label{sec:dxa-method}

The defect structure of the segregated and annealed-random states was compared at matched strain
on all three microstructures at the reference condition. Configurations were taken from the
deformation trajectories at the first recorded frame at or beyond \(\varepsilon = 0\), 0.02, 0.03,
0.045, 0.06, 0.08, 0.10, 0.15 and 0.20. For each frame the fcc, hcp and unclassified fractions were
obtained by a-CNA \cite{ref44}, and dislocation lines and their Burgers vectors by the dislocation
extraction algorithm (DXA) \cite{ref45}. Strain partitioning was measured with the atomic von Mises
shear strain \cite{ref51} and the non-affine squared displacement \(D^2_\text{min}\) \cite{ref49}, each computed
relative to the \(\varepsilon = 0\) frame of the same run with a 3.2 Å cutoff and averaged separately
over atoms that were unclassified (boundary) and fcc (grain) in that reference frame. These analyses
use the thermal configurations as recorded; segregated and annealed states were analysed at the same
temperature with identical settings, and only their ratio is interpreted. The random arm was not
included because its trajectories for microstructures 2 and 3 were not retained.

\subsection{Segregation-energy calculations}\label{auxiliary-atomistic-calculations}

Single-solute segregation energies were computed over the boundary network of microstructure 1 by a
two-region embedded-cluster method in the spirit of the polycrystalline segregation-energy spectrum
of Wagih and Schuh \cite{ref31,ref32}. For each of \(\approx 1500\) candidate boundary sites, a sphere
of radius 25 Å was carved from the relaxed polycrystal; atoms within 15 Å were relaxed and the
surrounding 10 Å shell, thicker than the potential cutoff, was held fixed, with and without a single
substitutional solute at the central site, giving
\[
E_\text{seg}^X = \left(E^\text{GB}_{X} - E^\text{GB}_{0}\right)
- \left(E^\text{bulk}_{X} - E^\text{bulk}_{0}\right),
\tag{7}
\]
with \(E_\text{seg}^X < 0\) denoting boundary binding. Sites at which solute insertion triggered a
local rearrangement were identified by the maximum relaxation displacement in excess of the bulk
reference value and censored above a cut \(d_\text{exc}\), which was swept from 0.02 to 2.0 Å.
Because the fraction of strongly binding sites depends on this cut, such fractions are reported as
ranges over it. The cluster radius was not varied, so \(E_\text{seg}\) is used comparatively, for its
sign and for nickel against chromium, rather than as a converged absolute value. The same
dilute single-solute calculation was carried out at symmetric \(\Sigma 5\,(310)\) and
\(\Sigma 13\,(320)\) [001] tilt bicrystals (Supplementary Section~S7).

\subsection{Statistics}\label{statistics-and-the-matched-control-requirement}

All comparisons use \(n = 3\) independent microstructures. Because each segregated run and its
controls share a tessellation, every comparison is paired within microstructure; two-sided paired
\(t\)-tests on the lives are reported together with the number of microstructures showing the effect
in the same direction. The chemistry contribution across all fifteen segregated--annealed pairs of
the condition and stress series was additionally estimated with a linear mixed model in which the
test condition is a fixed effect and the microstructure a random effect; for this balanced design the
test of the mean has two degrees of freedom. Replicate independence was verified for every series:
repeats that differ only in thermostat seed on a single tessellation reproduce each other roughly
five times more closely than independent tessellations do, and would understate the uncertainty
accordingly.

\section{Results}\label{results}

\subsection{Segregation redistributes all three species, each to a distinct position}\label{sec:seg}

\begin{figure}[pos=htbp]\centering
  \includegraphics{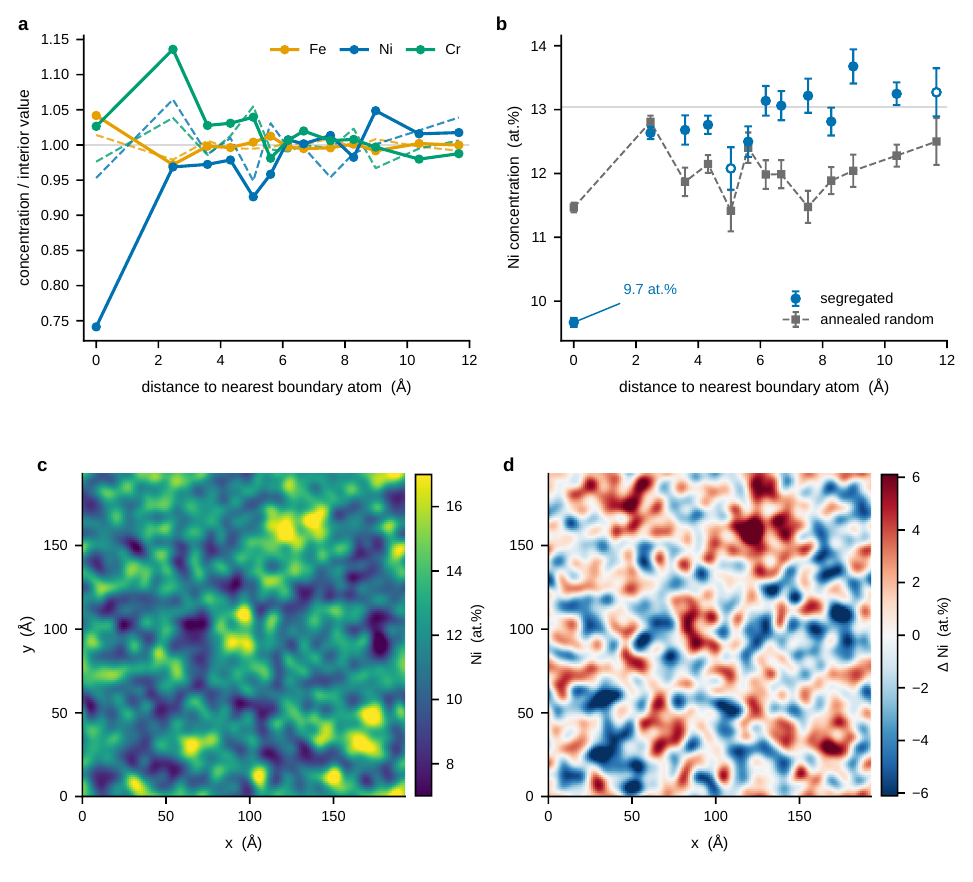}
  \caption{Grain-boundary composition after Monte Carlo annealing (microstructure 1). (a)~Shell-resolved
  concentration of Fe, Ni and Cr as a function of distance to the nearest boundary atom, normalised to
  the far-field interior, for the segregated state (solid lines, filled circles) and its
  annealed-random control (dashed lines); nickel depletes at the boundary plane, chromium enriches in
  the adjacent first fcc shell. (b)~Absolute nickel concentration versus the same distance for both
  states: 9.7~at.\% at the boundary plane against 13.0~at.\% in the interior, a boundary/interior
  ratio of $0.74$. Open symbols mark shells containing fewer than $10^4$ atoms. (c)~Nickel
  concentration on a representative mid-plane slab of the segregated state. (d)~The
  segregated-minus-annealed nickel difference on that slab, RMS $\Delta = 2.72$~at.\%.}
  \label{fig:profile}
\end{figure}

Hybrid Monte Carlo annealing at 900 K moves all three species relative to the grain boundaries, and
it moves them to different places (Fig.~\ref{fig:profile}). Measured as the segregated-minus-annealed
contrast of Eq.~(3) over the three microstructures, nickel is depleted at the boundary plane, by
\(\Delta R_\text{Ni} = -0.21 \pm 0.01\) (3/3 microstructures), falling to \(-0.08 \pm 0.03\) in the
first fcc shell. Chromium is enriched in the first fcc shell, by \(\Delta R_\text{Cr} = +0.09 \pm 0.02\)
(3/3), against \(+0.04 \pm 0.02\) at the plane itself, and iron is enriched at the plane by
\(+0.030 \pm 0.002\) (3/3). In absolute terms the segregated state gives a boundary-plane nickel ratio
of \(0.744 \pm 0.006\) against \(0.949 \pm 0.008\) for the annealed control, and a first-shell chromium
ratio of \(1.127 \pm 0.010\) against \(1.039 \pm 0.013\). Every contrast is confined to the boundary
plane and its first neighbour shell; the grain interiors remain at the nominal composition, and the
boundary fraction is unchanged by segregation.

The chromium enrichment depends on the reference used to measure it. Referenced to the fcc
population as a whole rather than to the far field, the chromium contrast is
\(+0.010 \pm 0.014\), consistent in sign in only two of three microstructures and indistinguishable
from zero. Chromium enriches in the first fcc shell, which belongs to the fcc population; a partition
reference places the chromium-rich material in the denominator and cancels the signal. Measured
against the far field, the enrichment keeps its sign in all three microstructures at all fifteen
cut combinations (Supplementary Table~S1). The nickel depletion, by contrast, is resolved under
either reference, because it is located on the boundary plane itself.

\subsection{Constant-stress response and time to 50\% strain}\label{sec:life}

\begin{figure}[pos=htbp]\centering
  \includegraphics{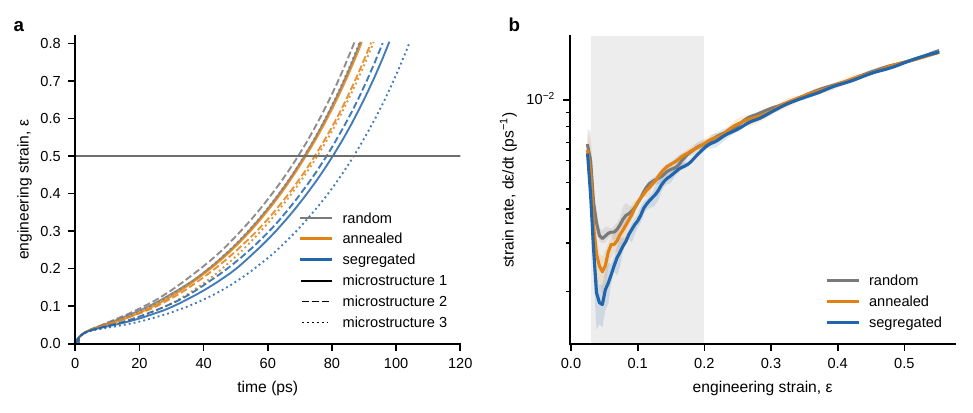}
  \caption{Constant-stress deformation at 300~K and 4.6~GPa. (a)~Engineering strain against time for
  the random, annealed-random and segregated states of the three microstructures; the horizontal line
  marks $\varepsilon = 0.5$, which defines $t_{0.5}$. (b)~Strain rate against strain, obtained from the
  smoothed histories of (a); lines are the mean of the three microstructures and bands their range.
  The shaded strain interval, $\varepsilon = 0.03$--$0.20$, is the one over which the integrated
  suppression $s_\mathrm{int}$ is evaluated.}
  \label{fig:curves}
\end{figure}

Fig.~\ref{fig:curves}a shows the strain histories of all nine runs at 300 K and 4.6 GPa. Within
2 ps of load application every specimen reaches \(\varepsilon = 0.023\)--0.024, independent of state,
after which the strain rate falls to a minimum at \(\varepsilon = 0.03\)--0.07 and then rises
steadily; no steady-state stage is present. The three arms separate at the rate minimum: the
segregated specimens reach the lowest minimum rate and the random specimens the highest
(Fig.~\ref{fig:curves}b), and the curves converge again beyond \(\varepsilon \approx 0.2\).

\begin{figure}[pos=htbp]\centering
  \includegraphics{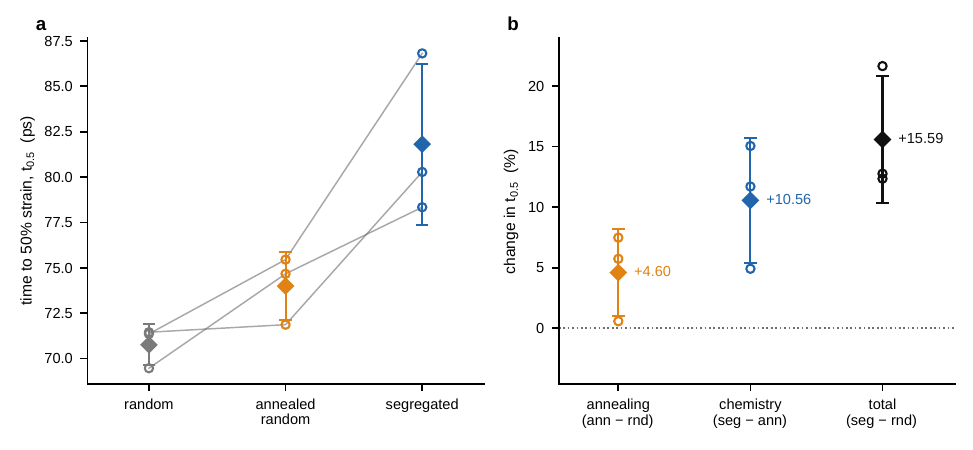}
  \caption{Time to 50\% strain at 300~K / 4.6~GPa, $n = 3$ microstructures. (a)~$t_{0.5}$ of the random,
  annealed-random and segregated states; open symbols are the individual microstructures, joined by
  grey lines within each microstructure, filled diamonds the mean $\pm$ SD. (b)~The annealing
  (annealed~$-$~random), chemistry (segregated~$-$~annealed) and total (segregated~$-$~random)
  contributions, labelled with their means; the corresponding paired tests are given in
  Table~\ref{tab:life}.}
  \label{fig:life}
\end{figure}

Table~\ref{tab:life} gives \(t_{0.5}\) for all three arms and microstructures. The chemistry
contribution is \(+10.56 \pm 5.17\)\%, the annealing contribution \(+4.60 \pm 3.59\)\%, and the total
segregated-versus-random effect \(+15.59 \pm 5.25\)\% (mean \(\pm\) sample SD, \(n = 3\);
Fig.~\ref{fig:life}). The total is resolved at this sample size (paired \(p = 0.037\); 95\% CI
\(+2.5\) to \(+28.6\)\%), whereas the chemistry (\(p = 0.073\)) and annealing (\(p = 0.155\))
contributions are positive in all three microstructures but individually have confidence intervals
that include zero. Within each microstructure the two contributions combine multiplicatively; applied to the
means, \(1.1056 \times 1.0460 - 1 \approx 15.6\)\%. The chemistry contribution is resolved when all fifteen
segregated--annealed pairs of the condition and stress series (Section~\ref{sec:ts}) are pooled:
the mixed model gives a mean gain of \(+7.4\)\% (95\% CI \(+0.8\) to \(+14.1\)\%, \(p = 0.040\),
two degrees of freedom), and all fifteen pairs are positive. With only three independent
microstructures this test has low power, and the result is best read together with the sign
consistency rather than as a confirmatory test on its own. Because the gain decreases from one
condition to the next, this pooled value is an average over the conditions sampled rather than an
estimate for any single condition. The \(t_{0.5}\) values use species labels assigned at construction
and interpolated strain histories only, so they are independent of the structural classification of
Section~\ref{grain-boundary-definition-and-the-composition-estimator}.

\begin{table}[pos=htbp]
\caption{Time to 50\% strain at 300~K / 4.6~GPa and its decomposition. Chemistry = segregated vs
annealed-random; annealing = annealed-random vs random; total = segregated vs random. Percentages are
computed from the unrounded lives; $p$ values are two-sided paired $t$-tests on the lives.}\label{tab:life}
\centering\small
\begin{tabular}{@{}lcccccc@{}}
\toprule
 & \multicolumn{3}{c}{\(t_{0.5}\) (ps)} & \multicolumn{3}{c}{change in \(t_{0.5}\) (\%)} \\
\cmidrule(lr){2-4}\cmidrule(l){5-7}
microstructure & random & annealed & segregated & chemistry & annealing & total \\
\midrule
1 & 71.45 & 71.87 & 80.28 & +11.71 & +0.58 & +12.35 \\
2 & 69.47 & 74.67 & 78.34 & +4.91 & +7.48 & +12.77 \\
3 & 71.36 & 75.46 & 86.81 & +15.05 & +5.73 & +21.65 \\
mean \(\pm\) SD & 70.76 \(\pm\) 1.12 & 74.00 \(\pm\) 1.89 & 81.81 \(\pm\) 4.44 & +10.56 \(\pm\) 5.17 & +4.60 \(\pm\) 3.59 & +15.59 \(\pm\) 5.25 \\
paired \(p\) & --- & --- & --- & 0.073 & 0.155 & 0.037 \\
95\% CI & --- & --- & --- & [\(-2.3\), \(+23.4\)] & [\(-4.3\), \(+13.5\)] & [\(+2.5\), \(+28.6\)] \\
\bottomrule
\end{tabular}
\end{table}

\subsection{The difference originates below a strain of 0.2}\label{sec:band}

\begin{figure}[pos=htbp]\centering
  \includegraphics{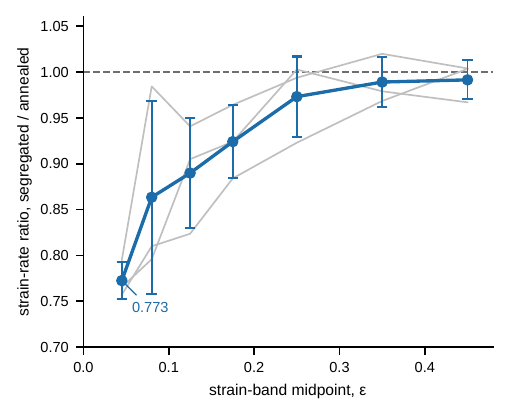}
  \caption{Ratio of segregated to annealed-control strain rate in successive strain bands at
  300~K / 4.6~GPa. Thin grey traces are the three microstructures, the heavy trace their mean
  $\pm$ SD, and the dashed line marks parity. $n = 3$, paired within microstructure.}
  \label{fig:band}
\end{figure}

The band decomposition locates the difference (Fig.~\ref{fig:band}). In the first band,
\(\varepsilon = 0.03\)--0.06, the segregated strain rate is \(0.773 \pm 0.020\) of its annealed
control. The ratio then rises through 0.863, 0.890 and 0.924 in the next three bands and is 0.973,
0.989 and 0.991 beyond \(\varepsilon = 0.20\). All three microstructures lie below unity in every
band up to \(\varepsilon = 0.20\); beyond it, no band has all three below unity.

The minimum strain rate falls across the three arms in every microstructure; for
microstructure 1, \(2.74 \times 10^{-3}\) (random), \(2.44 \times 10^{-3}\) (annealed) and
\(1.74 \times 10^{-3}\) ps\(^{-1}\) (segregated). The Monkman--Grant product
\(\dot{\varepsilon}_\text{min}\,t_{0.5}\) is not conserved between arms: \(0.213 \pm 0.018\) for random,
\(0.170 \pm 0.029\) for annealed-random and \(0.140 \pm 0.024\) for segregated. With the fixed strain
endpoint, this means that the minimum rate falls further than the time-averaged rate, that is, the
suppression is concentrated around the rate minimum rather than spread over the whole history.

Across the twelve segregated--annealed pairs of the stress and condition series for which the band
decomposition is available (four conditions in triplicate), the gain in \(t_{0.5}\) follows the
integrated suppression \(s_\text{int}\) with a slope of 0.84 and \(r^2 = 0.98\). Because
\(t_{0.5} = \int_0^{0.5} \mathrm{d}\varepsilon/\dot{\varepsilon}\), this relation is largely a kinematic
consequence of the suppression being confined to \(\varepsilon < 0.2\) (Supplementary Section~S6). The
non-trivial result is that the suppression in the first band alone, which spans about one sixteenth
of the strain to the endpoint, predicts the gain with \(r^2 = 0.65\) across conditions that differ by
650 K and 1.3 GPa (\(p = 0.006\) on the four condition means). The eighteen pairs of the matched
series (Section~\ref{sec:ts}), which were not used in this fit, follow the same relation: predicted
from their first-band suppression, their gains have a mean residual of \(+0.7\) and a root-mean-square
residual of 2.1 percentage points, against 2.3 within the fitted set.

\subsection{Defect structure and strain partitioning during early flow}\label{sec:mech}

\begin{figure}[pos=htbp]\centering
  \includegraphics{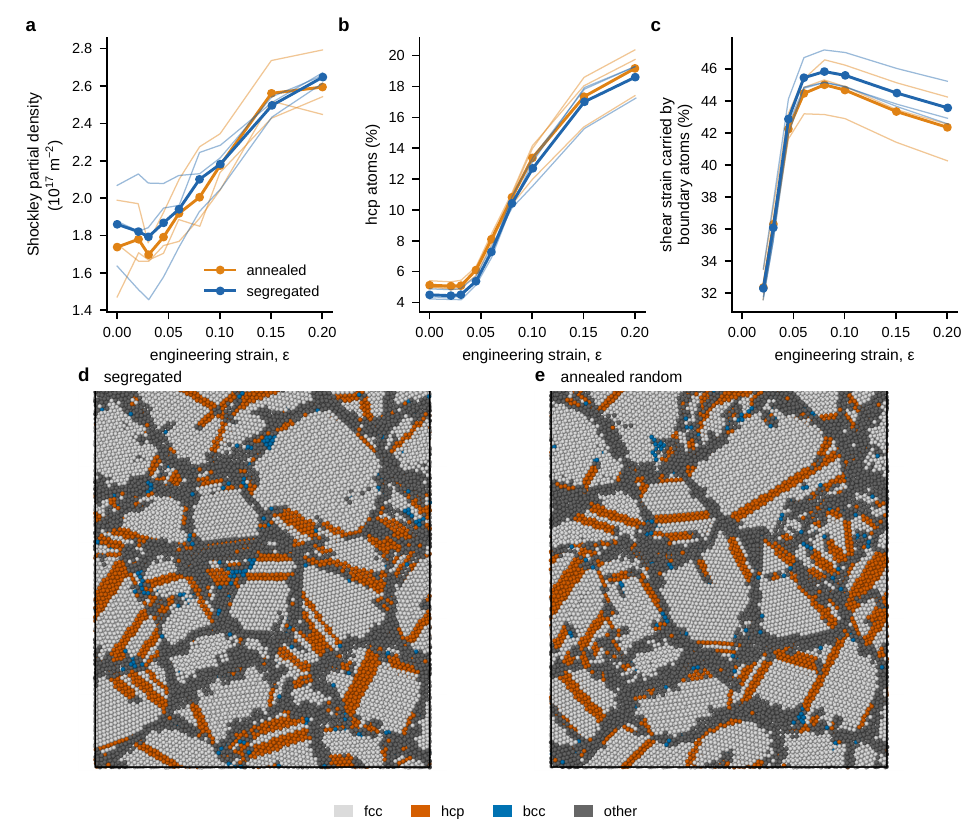}
  \caption{Defect structure and strain partitioning of the segregated and annealed-random states
  during deformation at 300~K / 4.6~GPa, against the strain of each analysed frame. (a)~Shockley
  partial dislocation density from DXA. (b)~Fraction of hcp atoms from adaptive CNA. (c)~Share of
  the total atomic shear strain carried by atoms that were unclassified (boundary atoms) at
  $\varepsilon = 0$. Thin lines are the three microstructures and heavy lines with symbols their mean.
  (d,~e)~1.5~nm mid-plane slabs of microstructure 1 at $\varepsilon = 0.10$, coloured by CNA, for the
  segregated and annealed-random states.}
  \label{fig:mechanism}
\end{figure}

Fig.~\ref{fig:mechanism} compares the segregated and annealed-random states at matched strain. The
density of Shockley partial dislocations rises from \(\approx 1.8 \times 10^{17}\) to
\(\approx 2.6 \times 10^{17}\) m\(^{-2}\) between \(\varepsilon = 0\) and 0.2 in both states, and the hcp
fraction, which records the stacking faults and twins left by partial slip, rises from about 5\% to
about 19\%. Over \(\varepsilon = 0.03\)--0.20 the segregated-to-annealed ratio, averaged over the
frames of each microstructure, is \(1.02 \pm 0.04\) for the total dislocation line density,
\(1.02 \pm 0.07\) for the Shockley partial density and \(1.06 \pm 0.12\) for perfect
\(\tfrac{1}{2}\langle 110\rangle\) dislocations; none is consistent in sign across the three
microstructures. The hcp fraction is lower in the segregated state in all three microstructures
(ratio \(0.93 \pm 0.04\)), but the difference is already present before loading
(\(0.88 \pm 0.10\) at \(\varepsilon = 0\)) and narrows with strain.

Strain partitioning differs slightly, and for most measures consistently. Over the same interval the mean shear strain
of boundary atoms is higher in the segregated state (ratio \(1.015 \pm 0.012\), 3/3), that of grain
atoms lower (\(0.983 \pm 0.013\), 3/3), and the non-affine displacement of boundary atoms higher
(\(1.033 \pm 0.021\), 3/3). At a given macroscopic strain the boundary atoms of the segregated
specimens therefore deform slightly more, and the grain atoms slightly less, than in the annealed
state; the share of the total shear strain carried by boundary atoms rises accordingly, by a ratio of
\(1.018 \pm 0.019\) (2/3; Fig.~\ref{fig:mechanism}c), without a resolvable change in dislocation content. No change in
the operating defect mechanism is detected: in both states plasticity proceeds by partial-dislocation
emission, faulting and twinning, at rates per unit strain that do not differ within the scatter of the
three microstructures (Fig.~\ref{fig:mechanism}d,e).

\subsection{Flow stress at constant strain rate}\label{sec:csr}

\begin{figure}[pos=htbp]\centering
  \includegraphics{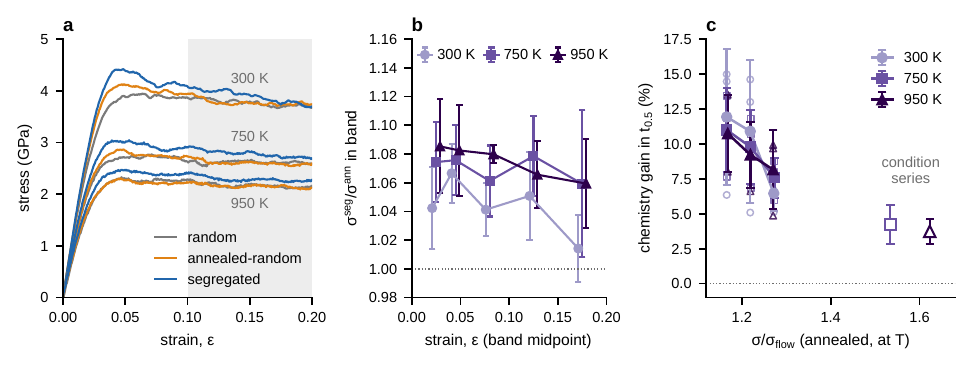}
  \caption{Strength at constant strain rate and the chemistry gain at matched relative stress; $n = 3$
  microstructures throughout. (a)~Stress--strain curves at $10^{9}$~s$^{-1}$ at 300, 750 and 950~K, mean of
  the three microstructures for each state; the shaded band marks the interval $\varepsilon = 0.10$--0.20
  over which $\sigma_\mathrm{flow}$ is averaged. (b)~Ratio of the mean stress of the segregated state to that
  of its annealed control in each strain band, mean $\pm$ SD. (c)~Chemistry gain in $t_{0.5}$ against the
  applied stress divided by the mean annealed flow stress at the test temperature. Small symbols are the
  individual microstructures and large symbols the mean $\pm$ SD at each level; the two isolated large
  open symbols are the 750~K/4.0~GPa and 950~K/3.5~GPa points of the condition series.}
  \label{fig:tsigma}
\end{figure}

Fig.~\ref{fig:tsigma}a shows the response at a constant strain rate of \(10^{9}\) s\(^{-1}\). In every
run the stress passes through a peak at \(\varepsilon = 0.04\)--0.14 and then softens gradually, and the
flow stress of the annealed controls falls from \(3.77 \pm 0.08\) GPa at 300 K to \(2.61 \pm 0.07\) and
\(2.16 \pm 0.07\) GPa at 750 and 950 K, by 31 and 43\%. At every temperature the segregated state is the
strongest of the three. Relative to its annealed control, its peak stress is higher by
\(6.3 \pm 2.9\), \(6.0 \pm 3.4\) and \(6.7 \pm 2.3\)\% at 300, 750 and 950 K (paired \(p = 0.056\), 0.085
and 0.029), and its mean stress over the first strain band, \(\varepsilon = 0.03\)--0.06, by
\(6.7 \pm 2.1\), \(7.6 \pm 2.4\) and \(8.2 \pm 3.2\)\% (\(p = 0.026\), 0.028 and 0.038), in all three
microstructures in each case; the flow stress rises by \(3.2 \pm 1.6\), \(6.9 \pm 3.9\) and
\(6.2 \pm 2.4\)\%. The annealing channel is smaller and less consistent: at 300 K it raises the
first-band stress by \(6.5 \pm 1.0\)\% (\(p = 0.009\)) but leaves the flow stress unchanged (\(+0.1\)\%),
and at 950 K it is not resolved. Against the random state, the segregated peak stress at 300 K is
higher by \(9.9 \pm 0.8\)\% (\(p = 0.001\)).

At 300 K the chemistry strengthening is concentrated early in flow (Fig.~\ref{fig:tsigma}b): the
ratio of segregated to annealed stress is \(1.067 \pm 0.021\) over \(\varepsilon = 0.03\)--0.06 and
\(1.014 \pm 0.023\) over 0.15--0.20, the strain range over which the constant-stress strain rate is most
suppressed and then recovers (Fig.~\ref{fig:band}). At 750 and 950 K the ratio stays between 1.06 and
1.09 up to \(\varepsilon = 0.20\). Because the stress softens after the peak in 25 of the 27 runs, by up
to 0.46 GPa between \(\varepsilon = 0.10\) and 0.20, the flow-stress contrast depends on the averaging
interval; the peak and first-band stresses are the more robust comparisons. The elastic modulus is also
slightly higher in the segregated state, by \(2.7 \pm 0.5\)\% at 300 K (\(p = 0.008\)). The individual
curves and per-run values are given in Supplementary Section~S9 (Fig.~S6, Table~S4).

\subsection{Temperature and stress dependence}\label{sec:ts}

Table~\ref{tab:series} collects the chemistry gain for the condition, stress and matched series,
with the applied stress expressed relative to the annealed flow stress at the test temperature. The
gain is positive in every microstructure at all eleven conditions, 33 of 33 paired comparisons, but
it is not constant (Fig.~\ref{fig:tsigma}c). In the condition series it is smaller at 750 K/4.0 GPa
(\(+4.22 \pm 1.41\)\%) and 950 K/3.5 GPa (\(+3.74 \pm 0.88\)\%) than at 300 K/4.6 GPa
(\(+10.90 \pm 5.10\)\%). Because the flow stress falls by 31 and 43\% over this temperature range
(Section~\ref{sec:csr}), those two tests lie at \(\sigma/\sigma_\text{flow} = 1.53\) and 1.62, well above
the 1.17--1.27 of the 300 K series, so temperature and relative stress are confounded in the condition
series. The largest effects are the least significant because microstructure 2 responds more weakly
than the other two, and does so most where the effect is largest.

\begin{table}[pos=htbp]
\caption{Chemistry gain in $t_{0.5}$, i.e.\ of the segregated state over its own-tessellation annealed
control, for the stress, condition and matched series. $\sigma/\sigma_\mathrm{flow}$ is the applied stress
over the mean annealed flow stress at the test temperature; $s_1$ is the first-band suppression of
Eq.~(6); $p$ values are two-sided paired $t$-tests on the lives. $^{\dagger}$The 4.6~GPa row uses the
original stress-series generation (Section~\ref{condition-stress-and-dose-series}; Supplementary Table~S2), for which the band decomposition
is not available; its mean gain agrees with the chemistry contribution of Table~\ref{tab:life}.}\label{tab:series}
\centering\small
\begin{tabular}{@{}ccccccc@{}}
\toprule
\(T\) (K) & \(\sigma\) (GPa) & \(\sigma/\sigma_\mathrm{flow}\) & gain per microstructure (\%) & mean \(\pm\) SD (\%) & paired \(p\) & \(s_1\) (\%) \\
\midrule
\multicolumn{7}{@{}l}{stress series} \\
300 & 4.4 & 1.17 & 14.99, 6.35, 14.46 & 11.93 \(\pm\) 4.84 & 0.055 & 27.45 \(\pm\) 1.60 \\
300 & 4.6 & 1.22 & 14.62, 5.08, 13.01 & 10.90 \(\pm\) 5.10 & 0.063 & n/a\(^{\dagger}\) \\
300 & 4.8 & 1.27 & 6.67, 5.12, 7.59 & 6.46 \(\pm\) 1.25 & 0.014 & 16.72 \(\pm\) 2.23 \\
\midrule
\multicolumn{7}{@{}l}{condition series} \\
750 & 4.0 & 1.53 & 5.25, 2.61, 4.80 & 4.22 \(\pm\) 1.41 & 0.035 & 9.47 \(\pm\) 1.22 \\
950 & 3.5 & 1.62 & 4.26, 2.72, 4.22 & 3.74 \(\pm\) 0.88 & 0.018 & 9.24 \(\pm\) 1.33 \\
\midrule
\multicolumn{7}{@{}l}{matched series} \\
750 & 3.04 & 1.17 & 11.89, 7.64, 13.45 & 10.99 \(\pm\) 3.00 & 0.027 & 23.46 \(\pm\) 3.79 \\
750 & 3.18 & 1.22 & 11.80, 6.82, 10.71 & 9.78 \(\pm\) 2.62 & 0.024 & 20.13 \(\pm\) 2.34 \\
750 & 3.32 & 1.27 & 8.78, 5.98, 7.98 & 7.58 \(\pm\) 1.44 & 0.012 & 17.99 \(\pm\) 0.87 \\
950 & 2.52 & 1.17 & 10.66, 8.02, 13.53 & 10.73 \(\pm\) 2.76 & 0.025 & 24.74 \(\pm\) 5.16 \\
950 & 2.63 & 1.22 & 9.88, 6.61, 11.15 & 9.22 \(\pm\) 2.34 & 0.024 & 21.64 \(\pm\) 2.01 \\
950 & 2.74 & 1.27 & 9.93, 4.87, 9.68 & 8.16 \(\pm\) 2.85 & 0.038 & 16.95 \(\pm\) 3.19 \\
\bottomrule
\end{tabular}
\end{table}

Within the 300 K stress series the gain decreases with increasing stress in all three
microstructures; the per-microstructure slopes are \(-20.8\), \(-3.1\) and \(-17.2\) percentage points
per GPa (mean \(-13.7 \pm 9.4\)). Between the 4.4 and 4.8 GPa endpoints the change is resolved in the
first-band suppression but not in the life gain itself (Table~\ref{tab:stress}), because the
microstructure-to-microstructure scatter of the rate metric is three to four times smaller.

\begin{table}[pos=htbp]
\caption{Change in each observable between the 4.4 and 4.8~GPa endpoints of the 300~K stress series
(4.8~GPa minus 4.4~GPa), with the two-sided paired $t$-test and the sign consistency across the three
microstructures.}\label{tab:stress}
\centering\small
\begin{tabular}{@{}lcccc@{}}
\toprule
tested quantity & \(\Delta\) per microstructure & mean \(\pm\) SD & paired \(p\) & consistent \\
\midrule
life gain (\%) & \(-8.3\), \(-1.2\), \(-6.9\) & \(-5.5 \pm 3.8\) & 0.127 & 3/3 \\
first-band suppression (pp) & \(-11.9\), \(-8.1\), \(-12.1\) & \(-10.7 \pm 2.2\) & 0.014 & 3/3 \\
integrated suppression (pp) & \(-7.9\), \(-1.5\), \(-9.6\) & \(-6.3 \pm 4.3\) & 0.125 & 3/3 \\
\bottomrule
\end{tabular}
\end{table}

The matched series removes this confounding. At the same \(\sigma/\sigma_\text{flow}\), the gains are
\(+11.0\), \(+9.8\) and \(+7.6\)\% at 750 K and \(+10.7\), \(+9.2\) and \(+8.2\)\% at 950 K over the three
levels, against \(+11.9\), \(+10.9\) and \(+6.5\)\% at 300 K, and each of the six matched conditions is
individually resolved (\(p = 0.012\)--0.038, 3/3; Table~\ref{tab:series}; per-run values in Supplementary
Table~S5). With the microstructure as the
unit of replication and the three levels averaged, the gain at 750 K differs from that at 300 K by
\(-0.3\) percentage points (95\% confidence interval \(-3.8\) to \(+3.2\), \(p = 0.74\)) and at 950 K by
\(-0.4\) (\(-4.0\) to \(+3.2\), \(p = 0.69\)). No dependence of the chemistry gain on temperature is
therefore resolved between 300 and 950 K, to within about \(\pm 4\) percentage points; the dependence
that remains is on \(\sigma/\sigma_\text{flow}\). The gain decreases with \(\sigma/\sigma_\text{flow}\) at
every temperature, but the per-microstructure slopes scatter too widely (at 300 K, \(-78\), \(-12\) and
\(-65\) percentage points per unit ratio) for a change of slope with temperature to be resolved. The
first-band suppression follows the gain: 23.5, 20.1 and 18.0\% at 750 K and 24.7, 21.6 and 16.9\% at
950 K over the three levels, against 27.4 and 16.7\% at the 300 K endpoints. The comparison does not
depend on the strength measure used for matching: the annealed peak stress falls with temperature by
the same fraction as the flow stress (to 0.70 and 0.56 of its 300 K value, against 0.69 and 0.57), so
matching on \(\sigma_\text{peak}\) would move the stress levels by less than 2\% and the temperature
differences by about 1 percentage point, leaving both unresolved (Supplementary Section~S11).

At 300 K, fitted within each arm, the life exponent is \(m = 2.68 \pm 0.28\) for the annealed controls and
\(3.25 \pm 0.55\) for the segregated specimens, larger in all three microstructures; the rate exponent
is larger and noisier, \(n = 7.0 \pm 3.4\) (annealed) and \(6.8 \pm 2.6\) (segregated; Supplementary
Table~S3). The two exponents differ because the Monkman--Grant product is not conserved. Over the
matched levels the annealed life exponent is \(m = 2.11 \pm 0.04\) at 750 K and \(1.95 \pm 0.18\) at
950 K. The quenched
boundary nickel ratio after the 950 K test is the same as after the 300 K test (Supplementary
Section~S8), as expected when substitutional diffusion is negligible on the \(10^2\) ps time scale
of the test; the elevated-temperature comparisons are therefore between the intended chemical
states.

\subsection{Site spectrum of segregation energies}\label{sec:eseg}

\begin{figure}[pos=htbp]\centering
  \includegraphics{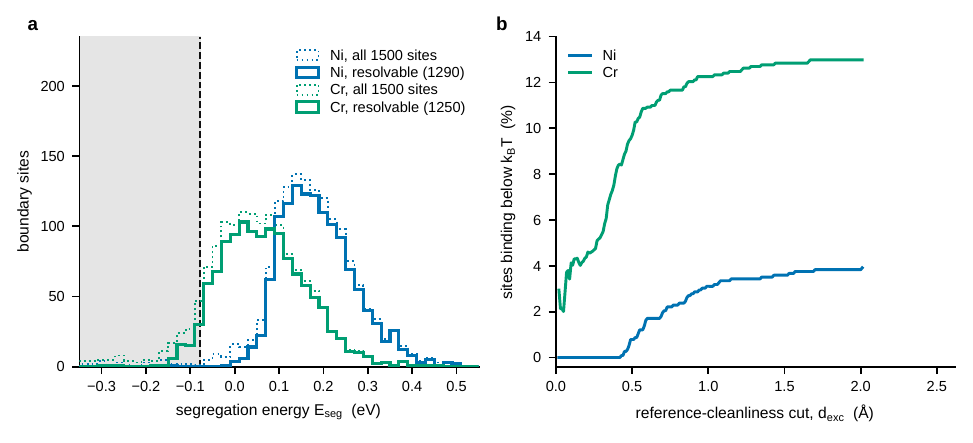}
  \caption{Single-solute grain-boundary segregation-energy site spectra over the boundary network
  (microstructure 1). (a)~Distribution of $E_\mathrm{seg}$ for Ni and Cr; dotted histograms are all
  sampled sites and solid histograms the subset resolvable without a large local rearrangement. The
  dashed line marks $-k_\mathrm{B}T$ at 900~K and the shaded region to its left is the binding side.
  (b)~Fraction of sites binding below $-k_\mathrm{B}T$ as a function of the rearrangement cut, which
  ranges over $0.0$--$3.9$\% for Ni and $2.0$--$13.0$\% for Cr.}
  \label{fig:eseg}
\end{figure}

Single-solute segregation energies over the boundary network (Fig.~\ref{fig:eseg}) show nickel to be
repelled from essentially every boundary site, while chromium binds at a minority of them. Over the
range of the rearrangement cut, 0.0--3.9\% of nickel sites and 2.0--13.0\% of chromium sites bind more
strongly than \(k_\text{B}T\) at 900 K. The ordering of the two species is independent of the cut and
matches the directions of Section~\ref{sec:seg}. Because these energies are computed one solute at a
time in the dilute limit, they involve no partition of the system into boundary and interior
populations. At the \(\Sigma 5\) and \(\Sigma 13\) tilt bicrystals nickel is likewise unfavourable at
the boundary plane, whereas chromium is near-neutral (Supplementary Fig.~S4).

\section{Discussion}\label{discussion}

\subsection{Segregation directions and the measurement of boundary composition}\label{sec:disc-seg}

The nickel depletion and chromium enrichment are consistent with high-throughput calculations of
grain-boundary segregation in the Fe--Ni--Cr system \cite{ref26} and with first-principles
segregation energies in \(\gamma\)-Fe, in which nickel is unfavourable at a subset of boundary sites
\cite{ref25}. They differ from the chromium-depleted, nickel-enriched boundaries characteristic of
irradiated austenitic steels, which are non-equilibrium steady states maintained by a point-defect
flux, and from the chromium depletion of sensitised steels, which is a consequence of carbide
precipitation \cite{ref7,ref8,ref9,ref41}. The carbon-free model alloy and the defect-free Monte Carlo
sampling exclude both routes by construction; the present calculation describes the
thermally annealed limit of the same alloy class. Enrichment located in the fcc shell adjacent to the boundary rather than on the
boundary plane has a precedent in hybrid MC/MD simulations of fcc multiprincipal element alloys
\cite{ref56}, and is the configuration that a structural partition of the system is least able to
detect.

The dependence of the chromium result on the reference region is the most transferable outcome of
this work. With a boundary fraction near 0.3, almost half of the nominally interior fcc atoms lie
within 5 Å of a boundary, and any solute that segregates there is measured partly against itself when
the boundary is referenced to the complementary structural population. The error grows with the
boundary fraction and is therefore largest in the nanocrystalline regime. The composition--classification
coupling identified by the label-permutation test is a second, independent reason to report
segregated-minus-annealed contrasts rather than absolute ratios: in a chemically homogeneous control
the boundary-plane nickel ratio is 0.95, not 1, and an absolute ratio would overstate the nickel
depletion by about a quarter.

\subsection{Origin of the early-strain suppression}\label{sec:disc-mech}

Under constant stress, a lower strain rate is the counterpart of a higher flow stress at fixed rate,
and the constant-strain-rate tests show this strengthening directly: the segregated state has a
6--7\% higher peak stress than its annealed control at all three temperatures, and at 300 K the
strengthening is largest over the strain band in which the constant-stress rate is most suppressed
(Section~\ref{sec:csr}). The suppression is located at \(\varepsilon = 0.03\)--0.2, just
after the initial transient and before the rate rises, and is accompanied by no resolvable change in
dislocation content and by a slightly larger share of strain carried by the boundaries. The data are
therefore consistent with a boundary contribution to the resistance to the onset of general plastic
flow: in the segregated state the boundary network accommodates marginally more strain without
additional dislocation activity, and the grains reach the same defect content at the same
macroscopic strain but later in time, because the strain rate is lower over this range
(Section~\ref{sec:band}). Two contributions to this resistance can be distinguished by
the controls. The annealing channel, positive in all microstructures, reflects structural relaxation
of the as-constructed boundaries, which is known to strengthen nanocrystalline alloys \cite{ref50}; the chemistry channel adds to it on the same
microstructures. Whether chemistry acts through the boundaries' resistance to sliding, through
dislocation emission from boundary sources, or through the lower pre-existing fault density found in
the segregated state cannot be separated with the present network-averaged analysis; a per-boundary correlation between
composition and sliding would do so. The size of the strengthening is consistent with the size of the
delay. Propagated through the annealed life exponent, as \((\sigma^\text{seg}/\sigma^\text{ann})^{m} - 1\),
the measured contrast predicts life gains of 9--18\%, depending on the strength measure and the
temperature, against 9--11\% measured at the middle stress level. It does not, however, reproduce the
microstructure-to-microstructure pattern of the gain (\(r = 0.25\) and 0.42 for the flow and peak
stress over the nine microstructure--temperature combinations; Supplementary Section~S11 and
Table~S6).

\subsection{Dependence on temperature and stress}\label{sec:disc-t}

The chemistry gain depends on the applied stress relative to the flow stress and, within the
resolution of three microstructures, not on temperature. The smaller gains of the condition series at
750 and 950 K arise because those tests were run at \(\sigma/\sigma_\text{flow} = 1.53\) and 1.62:
between 300 and 950 K the flow stress of the model falls by 43\%, while the chemistry strengthening at
fixed strain rate stays at 6--7\%. Both observations are consistent with a boundary resistance that
matters most near the onset of plastic flow and is overridden at a larger relative driving force; at
fixed relative stress there is no evidence that temperature assists the processes that override it.
Substitutional diffusion is negligible on the time scale of the tests, and the quenched boundary
composition after the 950 K test equals that after the 300 K test (Supplementary Section~S8), so the
segregated state is intact in every comparison. Comparing constant-stress responses across
temperature in such simulations therefore requires the applied stress to be referred to a separately
measured flow stress; a comparison at fixed applied stress, or at comparable \(t_{0.5}\) as in the
condition series, confounds temperature with relative stress. The dose series (Supplementary
Section~S2), a single realisation per Monte
Carlo temperature, gives a non-monotonic mechanical response and is not used in this interpretation.

All comparisons in this work are made within one interatomic model and at molecular-dynamics
stresses and time scales. Strain rates of \(10^{9}\)--\(10^{10}\) s\(^{-1}\) exceed laboratory values by
many orders of magnitude but are typical of atomistic studies of nanocrystalline deformation
\cite{ref2,ref57}; because the states of a microstructure are loaded identically and compared in
pairs, the conclusions concern differences between chemically distinct states rather than absolute
lives or strengths. The potential reproduces the lattice parameter and \(C_{11}\) of austenitic
stainless steel to within 3\% and places the model in the low-stacking-fault-energy regime
(Table~\ref{tab:validation}), but its solute--boundary energetics have not been compared with
first-principles segregation energies, and the segregated state is not converged in the magnitude
of segregation (Section~\ref{hybrid-molecular-dynamics-monte-carlo-segregation}); the segregation
directions are established, whereas their magnitudes are lower bounds. The mechanical statistics rest
on three microstructures, one of which responds systematically more weakly than the other two
(Section~\ref{sec:ts}).

\section{Conclusions}\label{conclusions}

Hybrid molecular-dynamics/Monte Carlo segregation of three nanocrystalline Fe--18Cr--12Ni
polycrystals, followed by constant-stress deformation of the same microstructures against annealed
and random controls, leads to the following conclusions.

\begin{enumerate}
\def\labelenumi{\arabic{enumi}.}
\item
  Monte Carlo annealing at 900 K depletes nickel at the grain-boundary plane and enriches chromium in
  the adjacent atomic shell; both directions are established at the production sampling budget and
  strengthen with further sampling of the one microstructure tested.
\item
  The chromium enrichment is resolved only when boundary composition is referenced to the far-field
  grain interior. The non-fcc/fcc partition places the enriched shell in its own reference and cancels
  the signal, an error that grows with the boundary fraction.
\item
  At 300 K and 4.6 GPa the segregated polycrystals take about 16\% longer than random ones to reach
  50\% strain. The difference arises from a lower strain rate in the early part of plastic flow, and
  the rate suppression in the first strain band alone accounts for about two-thirds of the variance
  of the life difference across the four test conditions (\(r^2 = 0.65\)) and predicts the eighteen
  pairs of the matched series, not used in the fit, with a root-mean-square residual of 2.1 percentage
  points.
\item
  At matched strain no difference in dislocation content between the segregated and annealed states
  is resolved, while the segregated boundaries carry a slightly larger share of the strain; no change in
  the defect mechanism of plasticity is detected.
\item
  At constant strain rate the segregated state has a 6--7\% higher peak stress than its annealed
  control at 300, 750 and 950 K. At matched ratios of applied to flow stress the gain in \(t_{0.5}\) is
  the same at all three temperatures, to within about \(\pm 4\) percentage points, and decreases as that
  ratio rises, consistent with a boundary resistance that is effective near the onset of plastic flow
  and is overridden at higher relative stress.
\end{enumerate}

The early part of a constant-stress test therefore carries most of the information on how boundary
chemistry affects deformation in this system, and is a more economical observable than the full
strain history. The natural next step is site-resolved: correlating per-boundary composition with
per-boundary sliding would identify the boundary population through which the chemistry acts.
\section*{Funding}
This research did not receive any specific grant from funding agencies in the public, commercial, or
not-for-profit sectors.

\section*{Declaration of competing interest}
The author declares that he has no known competing financial interests or personal relationships that
could have appeared to influence the work reported in this paper.

\section*{Declaration of generative AI and AI-assisted technologies in the manuscript preparation process}
During the preparation of this work the author used generative AI tools (Claude, Anthropic) in order to improve the language of the manuscript and to help with corrections of the analysis code.

\section*{Data availability}
The processed data underlying every figure and table, the loading histories of all deformation runs, and
the Python scripts used for the analysis and the figures are openly available on Zenodo \cite{ref62},
\url{https://doi.org/10.5281/zenodo.23189101}. All simulation parameters are given in Section~2, Table~1 and
Supplementary Table~S2. The raw simulation output (approximately 400 GB) is available from the author on
reasonable request.

\printcredits

\end{document}